\documentclass[prd,showpacs,preprintnumbers,amsmath,amssymb,10pt]{revtex4}

\usepackage{graphics}
\usepackage{epsfig}
\usepackage{subfigure}
\usepackage{dcolumn}
\usepackage{bm}
\usepackage{color}

\usepackage{multirow}
\usepackage{ulem}
\usepackage[colorlinks,
            citecolor=blue,
            anchorcolor=red,
            menucolor=red,
            linkcolor=red,
            filecolor=red,
            runcolor=red,
            urlcolor=blue,
            frenchlinks=red]{hyperref}
\begin{document}

\title{ Production of $X_b$ via $\Upsilon(5S, 6S)$ radiative decays }

\author{Xiao-Yun Wang$^1$, Zu-Xin Cai$^1$, Gang Li$^1$\footnote{gli@qfnu.edu.cn}, Shi-Dong Liu$^1$\footnote{liusd@qfnu.edu.cn},  Chun-Sheng An$^2$\footnote{ancs@swu.edu.cn}, and Ju-Jun Xie$^{3,4,5}$ }

\affiliation{$^1$ School of Physics and Physical Engineering, Qufu Normal University, Qufu 273165, People's Republic of China }
\affiliation{$^2$School of Physical Science and Technology, Southwest University, Chongqing 400715, China}
\affiliation{$^3$Institute of Modern Physics, Chinese Academy of Sciences, Lanzhou 730000, China}
\affiliation{$^4$School of Nuclear Science and Technology, University of Chinese Academy of Sciences, Beijing 101408, China}
\affiliation{$^5$Southern Center for Nuclear-Science Theory (SCNT), Institute of Modern Physics, Chinese Academy of Sciences, Huizhou 516000, Guangdong Province, China}

\begin{abstract}
We investigate the production of  $X_b$ in the process $\Upsilon(5S,6S)\to \gamma X_b$, where $X_b$ is assumed to be a $B {\bar B}^*$ molecular state. Two kinds of meson loops of $B^{(*)}{\bar B}^{(*)}$ and $B_1^{\prime}{\bar B}^{(*)}$ were considered. To explore the rescattering mechanism, we calculated the relevant branching ratios using the effective Lagrangian based on the heavy quark symmetry.
The branching ratios for the $\Upsilon(5S\,,6S) \to \gamma X_b$ were found to be at the orders of $10^{-7} \sim 10^{-6}$. Such sizeable branching ratios might be accessible at BelleII, which would provide important clues to the inner structures of the exotic state $X_b$.
\end{abstract}

\date{\today}







\maketitle

\section{Introduction}
\label{sec:introduction}

In the past decades, many \textit{XYZ} states have been observed by experiments~\cite{ParticleDataGroup:2022pth}. Some of them cannot be accommodated in the conventional quark model as $Q\bar Q$ ($Q=c$, $b$) and thus become excellent candidates for
exotic states. In order to understand the nature of the \textit{XYZ} states, many studies on their productions and decays have been carried out (for recent reviews, see Refs.~\cite{Chen:2016qju,Chen:2016spr,Esposito:2016noz,Guo:2017jvc,Olsen:2017bmm,Liu:2019zoy,Brambilla:2019esw,Guo:2019twa}). In 2003, the Belle Collaboration discovered an exotic candidate $X(3872)$ (also known as $\chi_{c1}(3872)$) in $B^+\to K^++ J/\psi \pi^+\pi^-$ decay~\cite{Choi:2003ue}. Subsequently, the $X(3872)$ was confirmed by several other experiments~\cite{Aubert:2004ns,Abazov:2004kp,Aaltonen:2009vj,Chatrchyan:2013cld,Aaij:2013zoa}. Its quantum numbers were determined to be $I^G(J^{PC})=0^+(1^{++})$~\cite{LHCb:2015jfc}. The $X(3872)$ has two salient features: the very narrow total decay width
($\Gamma_X < 1.2$ MeV), when compared to the typical hadronic width, and the closeness of mass to the threshold of $D^0 \bar{D}^{*0}$ ($M_{X(3872)}-M_{D^0}-M_{D^{*0} }=(-0.12\pm0.24)$~MeV)~\cite{ParticleDataGroup:2022pth}. These two features suggest that the $X(3872)$ might be a $\bar{D}D^*$ molecular state~\cite{Tornqvist:2004qy,Hanhart:2007yq}.

A lot of theoretical effort has been made to understand the nature of $X(3872)$ since its initial observation. Naturally, it follows to look for the counterpart with $J^{PC}=1^{++}$ (denoted as $X_b$ hereafter) in the bottom sector. These two states, which are related by heavy quark symmetry, should have some universal properties. The search for $X_b$ could provide us the discrimination between a compact multiquark configuration and a loosely bound hadronic molecule configuration. Since the mass of $X_b$ is very heavy and its $J^{PC}$ are $1^{++}$, a direct discovery is unlikely at the current electron-positron collision facilities, though the $\Upsilon(5S,6S)$ radiative decays are possible in the Super KEKB~\cite{Aushev:2010bq}.  In Ref.~\cite{He:2014sqj}, a search for $X_b$ in the $\omega \Upsilon(1S)$ final states has been presented, but no significant signal is observed. The production of $X_b$ at the LHC and the Tevatron~\cite{Guo:2014sca,Guo:2013ufa} and other exotic states at hadron colliders~\cite{Bignamini:2009sk,Artoisenet:2009wk,Artoisenet:2010uu,Esposito:2013ada,Ali:2011qi,Ali:2013xba} have been extensively investigated. In the bottomonium system, the isospin is almost perfectly conserved, which may explain the escape of $X_b$ in the recent CMS search~\cite{Chatrchyan:2013mea}. As a result, the radiative decays and isospin conserving decays are of high priority in searching $X_b$~\cite{Li:2014uia,Li:2015uwa,Karliner:2014sja,Wu:2016dws}.
In Ref.~\cite{Li:2014uia}, we have studied the
radiative decays $X_b \to \gamma \Upsilon(nS)$ ($n=1, 2, 3$), with $X_b$
being a candidate for the $B{\bar B}^*$ molecular state, and the partial widths into $\gamma X_b$ were found to be about $1$ keV. In this work, we revisit the $X_b$ production in $\Upsilon(5S,6S) \to \gamma X_b$ using the nonrelativistic effective field theory (NREFT). As is well known, the intermediate meson loop (IML) transition is one of the important nonperturbative transition mechanisms~\cite{Lipkin:1986bi,Lipkin:1988tg,Moxhay:1988ri}. Moreover, the recent studies on the productions and decays of exotic states~\cite{Guo:2010ak,Wang:2013cya,Liu:2013vfa,Guo:2013zbw,Cleven:2013sq,Chen:2011pv,Li:2012as,Li:2013xia,Bondar:2011ev,Chen:2012yr} lead to global agreement with the experimental data. Hence, to investigate the process $\Upsilon(5S,6S) \to \gamma X_b$, we calculated the IML contributions from both the $S$- and $P$-wave intermediate bottomed mesons.

The rest of the paper is organized as follows. In Sec.~\ref{sec:formula},
we present the theoretical framework used in this work. Then in Sec.~\ref{sec:results} the numerical results are presented, and a brief summary is given in Sec.~\ref{sec:summary}.

\section{Theoretical Framework}      \label{sec:formula}
\subsection{Triangle diagrams}   \label{sec:triangle-diagrams}

\begin{figure}[hbt]
\begin{center}
\includegraphics[width=0.95\textwidth]{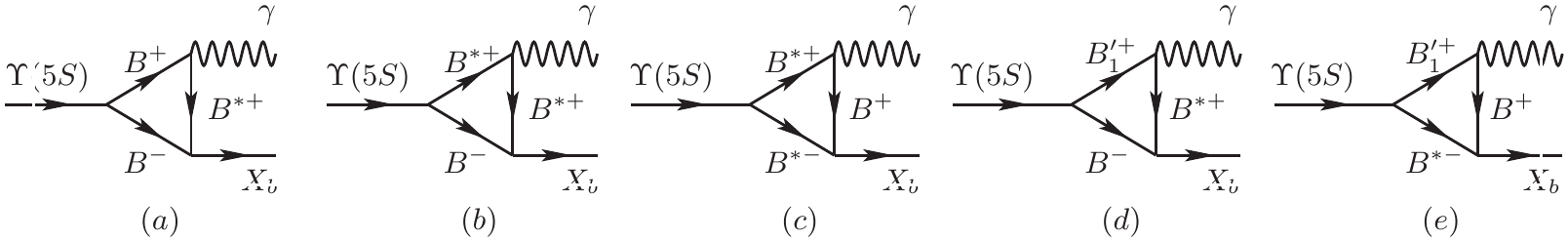}
\vglue-0mm\caption{Feynman  diagrams for $X_b$ production in $\Upsilon(5S) \to \gamma X_b  $ under the $B{\bar B}^*$ meson loop effects.} \label{fig:loops}
\end{center}
\end{figure}

Under the assumption that $X_b$ is a $B {\bar B}^*$ molecule, its production can be described by the triangle diagrams in Fig.~\ref{fig:loops}. With the quantum numbers of $1^{--}$, the initial bottomonium can couple to either two $S$-wave bottomed mesons in a $P$-wave, or one $P$-wave and one $S$-wave bottomed mesons in an $S$- or $D$-wave. The $X_b$ couples to the $B{\bar B}^*$ pair in an $S$-wave. Because the states considered here are close to the open bottomed mesons thresholds, the intermediate bottomed and antibottomed mesons are nonrelativistic. We are thus allowed to use a nonrelativistic power counting, the framework of which has been introduced to study the intermediate meson loop effects~\cite{Guo:2010ak}.
The three momentum scales as $v$, the kinetic energy scales as $v^2$, and each of the nonrelativistic propagator scales as $v^{-2}$. The $S$-wave vertices are independent of the velocity, while the $P$-wave vertices scales as $v$ or as the external momentum, depending on the process in question.

For the diagrams (a), (b), and (c) in Fig.~\ref{fig:loops}, the vertices involving the initial bottomonium are in a $P$-wave. The momentum in these vertices is contracted with the final photon momentum $q$ and thus should be counted as $q$. The vertices involving the photon are also in a $P$-wave, which should be counted as $q$. The decay amplitude scales as
\begin{equation}\label{eq:power-A}
\mathcal{A}_A\sim N_A\frac{v_A^5}{(v_A^2)^3}\frac{q^2}{m_B^2} = N_A \frac{E_\gamma^2}{v_A m_B^2} \,,
\end{equation}
where $E_\gamma$ is the external photon energy, $N_A$ contains all the constant factors, and $v_A$ is the average of the two velocities corresponding to the two cuts in the triangle diagram. While for the diagrams (d) and (e) in Fig.~\ref{fig:loops}, all the vertices are in $S$-wave. Then the amplitude for the Figs.~\ref{fig:loops}(d) and (e) scales as
\begin{equation}\label{eq:power-B}
\mathcal{A}_B\sim N_B\frac{v_B^5}{(v_B^2)^3}\frac{E_\gamma}{m_B}=N_B \frac{E_\gamma}{v_B m_B} \, .
\end{equation}

\subsection{Effective interaction Lagrangians}
\label{sec:effective Lagrangians}

To calculate the diagrams in Fig.~\ref{fig:loops}, we employ the effective Lagrangians constructed in the heavy quark limit. In this limit, the $S$-wave heavy-light mesons form a spin multiplet $H=(P,V)$ with $s_l^P=1/2^-$, where $P$ and $V$ denote the pseudoscalar and vector heavy mesons, respectively, i.e., $P(V)=(B^{(*)+},B^{(*)0},B_s^{(*)0})$. The $s_l^P=1/2^+$ states are collected in $S=(P_0^*,P_1^\prime)$ with $P_0^*$ and $P_1^\prime$ denoting the $B_0^*$ and $B_1^\prime$ states, respectively. In the two-component notation~\cite{Casalbuoni:1996pg,Hu:2005gf}, the spin multiplets
are given by
\begin{equation}\label{eq:HS-field}
    \begin{split}
        H_a &= \vec{V}_a\cdot\vec{\sigma}+P_a \,, \\
S_a &= \vec{P}_{1a}^\prime \cdot\vec{\sigma}+P_{0a}^* \,,
    \end{split}
\end{equation}
where $\vec \sigma$ is the Pauli matrix, and $a$ is the light flavor index. The fields for their charge conjugated mesons are
\begin{equation}\label{eq:HS-field-conjugated}
    \begin{split}
        {\bar H}_a &= -\vec{{\bar V}}_a\cdot\vec{\sigma}+ \vec {\bar P}_a \,, \\
        {\bar S}_a &= -\vec{{\bar P}}_{1a}^\prime \cdot\vec{\sigma}+{\vec {\bar P}}_{0a}^* \,.
    \end{split}
\end{equation}

Considering the parity, the charge conjugation, and the spin symmetry, the leading order Lagrangian for the coupling of the $S$-wave bottomonium fields to the bottomed and antibottomed mesons can be written as~\cite{Casalbuoni:1996pg}
\begin{equation}\label{eq:upsilon}
    {\cal L}_{\Upsilon(5S)} = i \frac{g_1}{2} Tr[{\bar H_a}^{\dagger} \vec{\sigma} \cdot \stackrel{\leftrightarrow}{\partial} H_a^{\dagger} \Upsilon] + g_2 Tr[{\bar H}_a^{\dagger} S_a^{\dagger} \Upsilon + {\bar S}_a^{\dagger} H_a^{\dagger} \Upsilon] + {\mbox {H.c.}}
\end{equation}
Here $A\stackrel{\leftrightarrow}{\partial} B=A(\partial B)-(\partial A)B$. The field for the $S$-wave $\Upsilon$ and $\eta_b$ is $\Upsilon= {\vec \Upsilon} \cdot {\vec \sigma} + \eta_b$. $g_1$ and $g_2$ are the coupling constants of $\Upsilon(5S)$ to a pair of $1/2^-$ bottom mesons and a $1/2^-$-$1/2^+$ pair of bottom mesons, respectively. We use $g_1^\prime$ and $g_2^\prime$ for the coupling constants of $\Upsilon(6S)$. Using the experimental branching ratios and widths of $\Upsilon(5S,6S)$~\cite{ParticleDataGroup:2022pth}, we get the coupling constants $g_1=0.1$ GeV$^{-3/2}$ and $g_1^\prime=0.08$ GeV$^{-3/2}$. On the other hand, we take $g_2=g_2^\prime = 0.05$ GeV$^{-1/2}$, as used in the previous work~\cite{Wu:2018xaa}.

To get the transition amplitude, we also need to know the photonic coupling to the bottomed mesons. The magnetic coupling of the photon to the $S$-wave bottomed mesons is described by the Lagrangian~\cite{Amundson:1992yp,Hu:2005gf}
\begin{equation}\label{eq:photon-HH}
    {\cal L}_{HH\gamma} = \frac{e\beta}{2} Tr[H_a^{\dagger} H_b \vec{\sigma} \cdot \vec{B} Q_{ab}] + \frac{e Q^\prime}{2 m_Q} Tr[H_a^{\dagger} \vec{\sigma} \cdot \vec{B} H_a] \,,
\end{equation}
where $Q= {\rm diag}\{2/3, -1/3, -1/3\}$ is the light quark charge matrix, and $Q^\prime$ is the heavy quark electric charge (in units of $e$). $\beta$ is an effective coupling constant and, in this work, we take $\beta\simeq 3.0$ GeV$^{-1}$, which is determined in the nonrelativistic constituent quark model and has been adopted in the study of radiative $D^*$ decays~\cite{Amundson:1992yp}. In Eq.~(\ref{eq:photon-HH}), the first term is the magnetic moment coupling of the light quarks, while the second one is the magnetic moment coupling of the heavy quark and hence is suppressed by $1/m_Q$. The radiative transition of the ${1/2}^+$ bottomed mesons to the ${1/2}^-$ states may be parameterized as~\cite{Colangelo:1993zq}
\begin{equation}
    \mathcal{L}_{SH\gamma} = -\frac{i e \widetilde{\beta}}{2} Tr[H_a^{\dagger} S_b \vec{\sigma} \cdot \vec{E} Q_{ba}] \, ,
\end{equation}
where $\widetilde{\beta}=0.42$ GeV$^{-1}$ is the same as used in Ref.~\cite{Mehen:2004uj}.

The $X_b$ is assumed to be an $S$-wave molecule with $J^{PC}=1^{++}$, which is given by the superposition of $B^0 {\bar B}^{*0}+c.c$ and $B^- {\bar B}^{*+}+c.c$ hadronic configurations:
\begin{eqnarray}
|X_b\rangle= \frac{1}{2} [(|B^0{\bar B}^{*0}\rangle - |B^{*0} {\bar B}^0\rangle) + (|B^+ B^{*-}\rangle - | B^- B^{*+}\rangle )].
\end{eqnarray}
Therefore, we can parameterize the coupling of $X_b$ to the bottomed mesons in terms of the following Lagrangian
\begin{equation}
{\cal L} = \frac{1}{2} X^{i \dagger} [x_1(B^{*0i} {\bar B}^{0} - B^0 {\bar B}^{*0i}) + x_2(B^{*+i} B^- - B^+ B^{*-i})] + {\mbox {H.c.}} \, ,
\end{equation}
where $x_i$ denotes the coupling constant. Since the $X_b$ is slightly below the $S$-wave $B{\bar B}^*$ threshold, the effective coupling of this
state is related to the probability of finding the $B{\bar B}^*$
component in the physical wave function of the bound
states and the binding energy, $\epsilon_{X_b}=m_B+m_{B^*}-m_{X_b}$~\cite{Weinberg:1965zz, Baru:2003qq,Guo:2013zbw}
\begin{equation}\label{eq:coupling-Xb}
x_i^2 \equiv 16\pi (m_B+ m_{B^*})^2 c_i^2 \sqrt{\frac {2\epsilon_{X_b}}{\mu}}\,,
\end{equation}
where $c_i=1/{\sqrt 2}$ and $\mu=m_B m_{B^*}/(m_B+m_{B^*})$ is the reduced mass. Here, it should be pointed out that the coupling constant $x_i$ in Eq.~(\ref{eq:coupling-Xb}) is based on the assumption that $X_b$ is a shallow bound state where the potential binding the mesons is short-ranged.

The decay amplitudes of the triangle diagrams in Fig.~\ref{fig:loops} can be obtained and the explicit
transition amplitudes for $\Upsilon(5S,6S) \to \gamma X_b$ are presented in Appendix~\ref{appendix}. The partial decay widths of $\Upsilon(5S,6S) \to \gamma X_b$ are given by
\begin{equation}\label{eq:Gamma3900}
\Gamma(\Upsilon(5S,6S) \to \gamma X_b) = \frac {E_\gamma  |{\cal M}_{\Upsilon(5S,6S) \to \gamma X_b}|^2 }{24\pi M_{\Upsilon(5S,6S)}^2}\, ,
\end{equation}
where $E_\gamma$ is the photon energies in the $\Upsilon(5S,6S)$ rest frame.

\section{Numerical Results}
\label{sec:results}


In Ref.~\cite{Du:2017zvv}, authors predicted a large width of 238 MeV for $B_1^\prime$. This large width effect for $B_1^\prime$ was taken into account in our calculations by using the Breit-Wigner (BW) parameterization to approximate the spectral function of the 1/2$^+$ bottom meson of width. The explicit formula for $B_1^{\prime}$ is
\begin{equation}
\mathcal{M}_{B_1^{\prime}} = \frac{1}{W_{B_1^{\prime}}} \int_{s_l}^{s_h} \mathrm{d}s \rho_{B_1^{\prime}}(s) \bar{\mathcal{M}}_{B_1^{\prime}} (s)\,,
\end{equation}
where $W_{B_1^{\prime}} = \int_{s_l}^{s_h} \mathrm{d}s\rho_{B_1^{\prime}}(s)$ is the normalization factor, $\bar{\mathcal{M}}_{B_1^{\prime}}(s)$ represents the loop amplitude of $B_1^{\prime}$ calculated using $s$ as the mass squared, $s_l=(M_B+m_{\gamma})^2$, $s_h=(M_{B_1^{\prime}}+\Gamma_{B_1^{\prime}})^2$, and $\rho_{B_1^{\prime}}(s)$ is the spectral function of $B_1^{\prime}$
\begin{equation}
\rho_{B_1^{\prime}}(s) = \frac{1}{\pi} \operatorname{Im}\frac{-1}{s - M_{B_1^{\prime}}^2+i M_{B_1^{\prime}} \Gamma_{B_1^{\prime}}}\,.
\end{equation}

Before proceeding to the numerical results, we first briefly review the predictions of the mass of $X_b$. The existence of the $X_b$ is predicted in both the tetraquark model~\cite{Ali:2009pi} and those involving a molecular interpretation~\cite{Tornqvist:1993ng,Guo:2013sya,Karliner:2013dqa}. In Ref.~\cite{Ali:2009pi}, the mass of the lowest-lying $1^{++}$ $\bar b \bar q bq$ tetraquark is predicated to be $10504$ MeV, while the mass of the $B\bar B^*$ molecular state is predicated to be a few tens of MeV higher~\cite{Tornqvist:1993ng,Guo:2013sya,Karliner:2013dqa}. For example, in Ref.~\cite{Tornqvist:1993ng}, the mass was predicted to be $10562$~MeV, corresponding to a binding energy of $42$ MeV, while with a binding energy of $(24^{+8}_{-9})$ MeV it was predicted to be $(10580^{+9}_{-8})$~MeV~\cite{Guo:2013sya}. Therefore, it might be a good approximation and might be applicable if the binding energy is less than $50$ MeV. In order to cover the range for the previous molecular and tetraquark predictions in Refs.~\cite{Ali:2009pi,Tornqvist:1993ng,Guo:2013sya,Karliner:2013dqa}, we performed the calculations up to a binding energy of $100$ MeV and choose several illustrative values of $\epsilon_{X_b} = (5,10,25,50,100)$ MeV for discussion.

In Table~\ref{table:upsilon5s}, we list the contributions of $\Upsilon(5S) \to \gamma X_b$ from $B^{(*)}{\bar B}^{(*)}$ loops, $B_1^{\prime}{\bar B}^{(*)}$ loops, and the total contributions. For the $B_1^\prime$, we choose the ${\Gamma}_{B_1^{\prime}}$ to be $0$, $100$ MeV and $200$ MeV, respectively. It can be seen that the contributions from $B^{(*)}{\bar B}^{(*)}$ loops are about $10^{-3}$ keV. For the contributions from $B_1^{\prime}{\bar B}^{(*)}$ loops, the partial decay widths decrease with increasing the width of $B_1^{\prime}$. Without the width effects of $B_1^\prime$, i.e., $\Gamma_{B^\prime}=0$, the contributions from  $B_1^{\prime}{\bar B}^{(*)}$ loops are about $10^{-2}$ keV, while with $\Gamma_{B_1^\prime}=200$ MeV the contributions are about two orders of magnitude smaller. As seen, the total decay widths also decrease with increasing the width of $B_1^{\prime}$. The obtained partial widths range from $10^{-3}$ to $10^{-2}$ keV, indicating a sizeable branching fraction  from about $10^{-7}$ to $10^{-6}$.

\begin{table}[tb]
\begin{center}
\caption{The predicted decay widths (in units of keV) of $\Upsilon(5S)\to\gamma X_b$ for different binding energies. Here we choose the ${\Gamma}_{B_1^{\prime}}$ to be $0$, $100$, and $200$ MeV, respectively. }\label{table:upsilon5s}
 \begin{tabular}{|l|c|c|c|c|c|c|c|}
 \hline
\multirow{2}{*}{Binding energy} & \multirow{2}{*}{$B^{(*)}{\bar B}^{(*)}$ loops} & \multicolumn{3}{c|}{$B_1^{\prime}{\bar B}^{(*)}$ loops} & \multicolumn{3}{c|}{Total Decay Widths}\\
\cline{3-8}
~ & ~ & $\Gamma_{B_1^{\prime}}=0$ & $\Gamma_{B_1^{\prime}}=100$ & $\Gamma_{B_1^{\prime}}=200$ & $\Gamma_{B_1^{\prime}}=0$ & $\Gamma_{B_1^{\prime}}=100$ & $\Gamma_{B_1^{\prime}}=200$ \\
\hline
 ${\epsilon}_{X_b}=5$ MeV & $7.24\times 10^{-4}$  & $2.22\times 10^{-2}$  & $1.43\times 10^{-3}$ & $3.13\times 10^{-4}$  & $2.77\times 10^{-2}$  & $3.25\times 10^{-3}$ & $1.49\times 10^{-3}$ \\ \hline
 ${\epsilon}_{X_b}=10$ MeV & $1.07\times 10^{-3}$  & $2.01\times 10^{-2}$  & $1.47\times 10^{-3}$ & $3.52\times 10^{-4}$  & $2.69\times 10^{-2}$  & $3.87\times 10^{-3}$ & $1.99\times 10^{-3}$  \\ \hline
 ${\epsilon}_{X_b}=25$ MeV & $1.92\times 10^{-3}$  & $1.55\times 10^{-2}$  & $1.41\times 10^{-3}$ & $3.95\times 10^{-4}$  & $2.41\times 10^{-2}$  & $5.00\times 10^{-3}$ & $3.05\times 10^{-3}$ \\ \hline
 ${\epsilon}_{X_b}=50$ MeV & $3.32\times 10^{-3}$  & $1.19\times 10^{-2}$  & $1.34\times 10^{-3}$ & $4.30\times 10^{-4}$  & $2.26\times 10^{-2}$  & $6.62\times 10^{-3}$ & $4.64\times 10^{-3}$  \\ \hline
 ${\epsilon}_{X_b}=100$ MeV & $6.80\times 10^{-3}$  & $9.48\times 10^{-3}$  & $1.34\times 10^{-3}$ & $4.91\times 10^{-4}$  & $2.49\times 10^{-2}$  & $1.05\times 10^{-2}$ & $8.39\times 10^{-3}$ \\ \hline
\end{tabular}
\end{center}
\end{table}

The results for $\Upsilon(6S) \to \gamma X_b$ are summarized in Table~\ref{table:upsilon6s}. The contributions from $B^{(*)}{\bar B}^{(*)}$ loops are about $10^{-3}$ keV. Different from the case of $\Upsilon(5S) \to \gamma X_b$, the contribution from $B_1^{\prime}\bar{B}^{(*)}$ loops for $\Upsilon(6S) \to \gamma X_b$ is not monotonous with the width of $B_1^{\prime}$. This finding indicate that the $B_1^{\prime}$ width has a smaller effect in $\Upsilon(6S) \to \gamma X_b$ than in $\Upsilon(5S) \to \gamma X_b$, which may be due to the fact that the mass of $\Upsilon(5S)$ is closer to the threshold of $B_1^{\prime}\bar{ B}^{(*)}$ than $\Upsilon(6S)$. It can be seen that the contributions from $B_1^{\prime}\bar{B}^{(*)}$ loops range from $10^{-4}$ to $10^{-3}$ keV, which is about $1$ order of magnitude smaller than $\Upsilon(5S)$. The total decay widths increase with increasing the width of $B_1^{\prime}$. Similar to the case of the process $\Upsilon(5S) \to \gamma X_b$ the obtained partial widths for $\Upsilon(6S) \to \gamma X_b$ are also about $10^{-3}$ to $10^{-2}$ keV, thereby corresponding to a branching fraction of about $10^{-7}$.

\begin{table}[htbp]
\begin{center}
\caption{The predicted decay widths (in units of keV) of $\Upsilon(6S)\to\gamma X_b$ for different binding energies. Here we choose the ${\Gamma}_{B_1^{\prime}}$ to be $0$, $100$, and $200$ MeV, respectively.  }
\label{table:upsilon6s}
 \begin{tabular}{|l|c|c|c|c|c|c|c|}
 \hline
\multirow{2}{*}{Binding energy} & \multirow{2}{*}{$B^{(*)}{\bar B}^{(*)}$ loops} & \multicolumn{3}{c|}{$B_1^{\prime}{\bar B}^{(*)}$ loops} & \multicolumn{3}{c|}{Total Decay Widths}\\
\cline{3-8}
~ & ~ & $\Gamma_{B_1^{\prime}}=0$ & $\Gamma_{B_1^{\prime}}=100$ & $\Gamma_{B_1^{\prime}}=200$ & $\Gamma_{B_1^{\prime}}=0$ & $\Gamma_{B_1^{\prime}}=100$ & $\Gamma_{B_1^{\prime}}=200$ \\
\hline
 ${\epsilon}_{X_b}=5$ MeV & $1.52\times 10^{-3}$  & $5.67\times 10^{-4}$  & $1.11\times 10^{-3}$ & $4.15\times 10^{-4}$  & $8.19\times 10^{-4}$  & $1.10\times 10^{-3}$ & $1.50\times 10^{-3}$ \\ \hline
 ${\epsilon}_{X_b}=10$ MeV & $2.22\times 10^{-3}$  & $7.52\times 10^{-4}$  & $1.27\times 10^{-3}$ & $5.11\times 10^{-4}$  & $1.25\times 10^{-3}$  & $1.62\times 10^{-3}$ & $2.20\times 10^{-3}$  \\ \hline
 ${\epsilon}_{X_b}=25$ MeV & $3.87\times 10^{-3}$  & $1.01\times 10^{-3}$  & $1.41\times 10^{-3}$ & $6.38\times 10^{-4}$  & $2.40\times 10^{-3}$  & $2.90\times 10^{-3}$ & $3.80\times 10^{-3}$ \\ \hline
 ${\epsilon}_{X_b}=50$ MeV & $6.39\times 10^{-3}$  & $1.17\times 10^{-3}$  & $1.45\times 10^{-3}$ & $7.27\times 10^{-4}$  & $4.39\times 10^{-3}$  & $4.99\times 10^{-3}$ & $6.19\times 10^{-3}$  \\ \hline
 ${\epsilon}_{X_b}=100$ MeV & $1.21\times 10^{-2}$  & $1.27\times 10^{-3}$  & $1.46\times 10^{-3}$ & $8.22\times 10^{-4}$  & $9.24\times 10^{-3}$  & $9.77\times 10^{-3}$ & $1.15\times 10^{-2}$ \\ \hline
\end{tabular}
\end{center}
\end{table}

\begin{figure}[tb]
	\centering
	\includegraphics[width=0.8\linewidth]{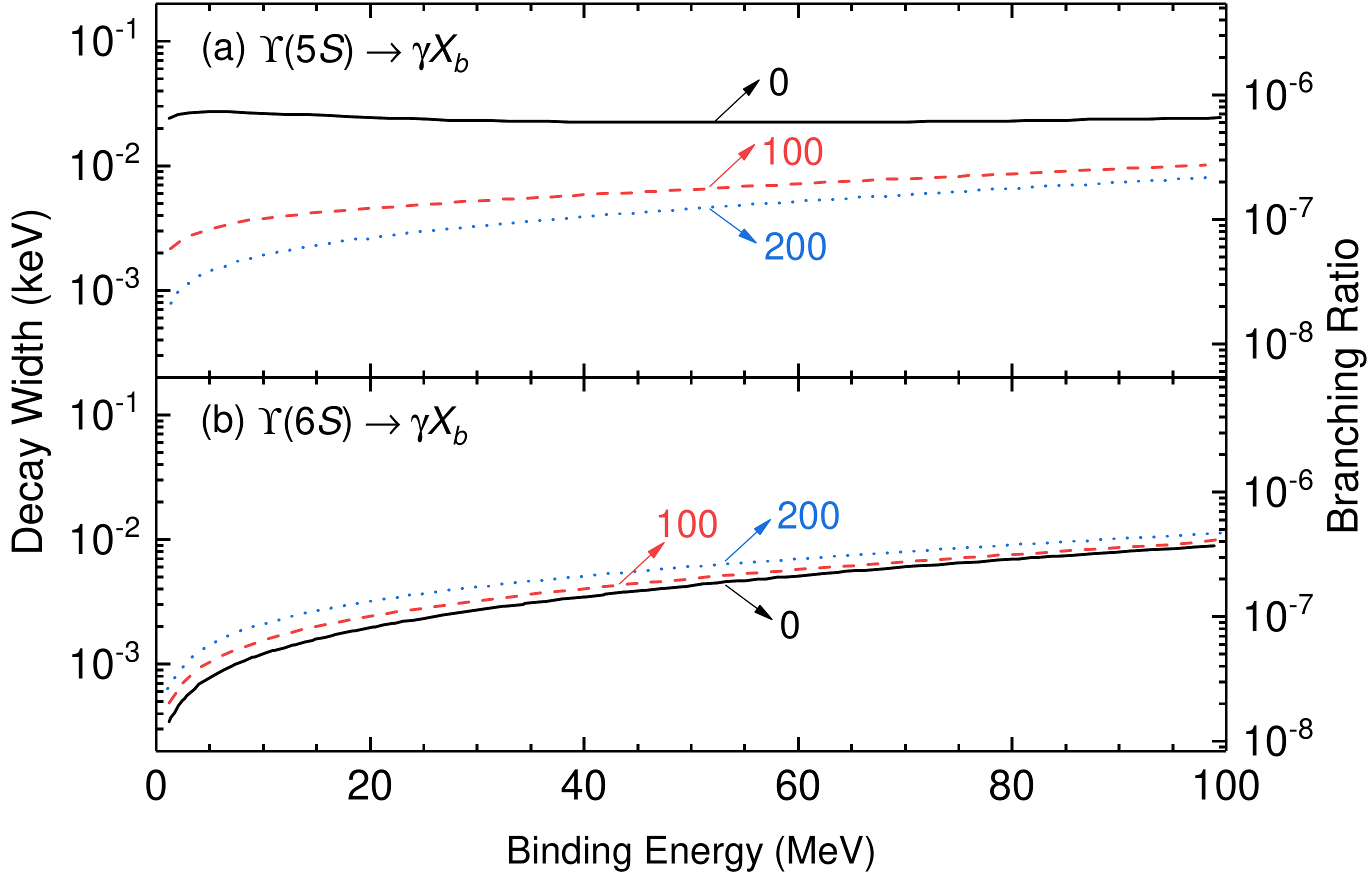}
	\caption{The dependence of the decay widths of $\Upsilon(5S)\to \gamma X_b$ (a) and $\Upsilon(6S)\to \gamma X_b$ (b) on the binding energy for different $B_1^{\prime}$ widths as indicated by the numbers in the graph. The right $y$-axis represents the corresponding branching ratio.}
	\label{fig:upsilon56s}
\end{figure}

In Fig.~\ref{fig:upsilon56s}(a), we plot the decay widths and the branching ratios of $\Upsilon(5S)\to\gamma X_b$ as a function of the binding energy with $\Gamma_{B_1^\prime} =0$ MeV (solid line), $\Gamma_{B_1^\prime} =100$ MeV (dash line), and $\Gamma_{B_1^\prime} =200$ MeV (dotted line). The coupling constants of $X_b$ in Eq.~(\ref{eq:coupling-Xb})
and the threshold effects can simultaneously influence the binding energy dependence of the partial widths. With increasing the binding energy $\epsilon_{X_b}$, the coupling strength of $X_b$ increases, and the threshold effects decrease. Both the coupling strength of $X_b$ and the threshold effects vary quickly in the small $\epsilon_{X_b}$ region and slowly in the large $\epsilon_{X_b}$ region. As a result, the partial width is relatively sensitive to the small $\epsilon_{X_b}$, while at the large $\epsilon_{X_b}$ region it keeps nearly constant. As seen, at the same binding energy, the partial widths with small $\Gamma_{B_1^\prime}$ are larger than those with large $\Gamma_{B_1^\prime}$

In Fig.~\ref{fig:upsilon56s}(b), the dependences of the decay widths and the branching ratios for $\Upsilon(6S)\to\gamma X_b$ on the binding energy are shown. Similar to the case of $\Upsilon(5S)\to \gamma X_b$, the partial width is relatively sensitive to the small $\epsilon_{X_b}$, while at the large $\epsilon_{X_b}$ region, it becomes nearly independent of the binding energy. As shown in this figure, at the same binding energy, the partial widths increases with the increase of $\Gamma_{B_1^\prime}$. It can be seen that the predicted partial width for $\Upsilon(6S)\to\gamma X_b$ is insensitive to the $B_1^\prime$ width, which is different from the case of $\Upsilon(5S)\to\gamma X_b$. This indicates that the intermediate bottomed meson loop contribution to the process $\Upsilon(6S)\to\gamma X_b$ is smaller than that to $\Upsilon(5S)\to\gamma X_b$.

\section{Summary}
\label{sec:summary}
We have presented the production of $X_b$ in the radiative decays of $\Upsilon(5S,6S)$. The $X_b$ is assumed to be a molecular state of $B \bar{B}^*$. The numerical calculations were performed under two kinds of intermediate bottomed meson loops. The first kind is $B^{(*)}{\bar B}^{(*)}$ loop coupled with $\Upsilon(5S,6S)$ in $P$-wave and the second is $B_1^{\prime}{\bar B}^{(*)}$ loop coupled with $\Upsilon(5S,6S)$ in $S$-wave. Our results show that the partial widths of $\Upsilon(5S\,,6S) \to \gamma X_b$ range from $10^{-3}$ to $10^{-2}$ keV, which correspond to the branching ratios from $10^{-7}$ to $10^{-6}$. In Refs.~\cite{Li:2014uia,Li:2015uwa}, we have studied the radiative decays and the hidden bottomonium decays of $X_b$. If we consider that the branching ratios of the isospin conserving process $X_b \to \omega \Upsilon(1S)$ are relatively large, a search for $\Upsilon(5S)\to \gamma X_b\to \gamma \omega \Upsilon(1S)$ may be possible for the updated BelleII experiments. These studies may help us investigate the $X_b$ deeply. The experimental observation of $X_b$ will provide us further insight into the spectroscopy of exotic states and is helpful to probe the structure of the states connected by the heavy quark symmetry.

\section*{Acknowledgements}
\label{sec:acknowledgements}

This work is partly supported by the National Natural Science Foundation of China under Grant Nos.
12075133, 12105153, 12075288, 11735003, 11961141012, and 11835015, and by the Natural Science
Foundation of Shandong Province under Grant Nos. ZR2021MA082, and ZR2022ZD26. It is also supported by Taishan
Scholar Project of Shandong Province (Grant No.tsqn202103062),
the Higher Educational Youth Innovation Science and Technology
Program Shandong Province (Grant No. 2020KJJ004), the Chongqing Natural Science Foundation under Project No. cstc2021jcyj-msxmX0078, and the Youth Innovation Promotion Association CAS.

\begin{appendix}
\section{The transition amplitudes}\label{appendix}


Here we give the amplitudes for the transitions $\Upsilon(5S,6S) \to \gamma X_b$. $\epsilon_1$, $\epsilon_2$, and $\epsilon_3$ are the polarization
vectors of the initial state $\Upsilon(5S,6S)$, final photon $\gamma$, and final state $X_b$, respectively. The transition amplitudes shown in Figs.~\ref{fig:loops} (a)-(c) are
\begin{eqnarray}
\mathcal{M}_a &=& -eg_1g_X \left(\beta Q + \frac{Q^{\prime}}{m_Q} \right) {\epsilon}_{ijk} q^i \epsilon_2^j \epsilon_3^k \epsilon_1 \cdot q I_a^{(1)}(m_B,m_B,m_{B^*},q)\,, \\
\mathcal{M}_b
&=& eg_1g_X \left(\beta Q - \frac{Q^{\prime}}{m_Q}\right) {\epsilon}_{ijk} \epsilon_1^i q^j (q\cdot \epsilon_3 \epsilon_2^k - q^k \epsilon_2\cdot \epsilon_3) I_b^{(1)}(m_{B^*},m_B,m_{B^*},q)\,, \\
\mathcal{M}_c &=& -eg_1g_X \left(\beta Q + \frac{Q^{\prime}}{m_Q} \right) {\epsilon}_{ijk} q^i \epsilon_2^j \left(\epsilon_1^k q\cdot \epsilon_3 - q\cdot\epsilon_1 \epsilon_3^k + q^k \epsilon_1 \cdot \epsilon_3 \right) I_c^{(1)}(m_{B^*},m_{B^*},m_B,q)\,.
\end{eqnarray}
The transition amplitudes shown in Figs.~\ref{fig:loops} (d) and (e) are
\begin{eqnarray}
\mathcal{M}_d &=& eQ \widetilde{\beta} g_2 g_X {\epsilon}^{ijk} \epsilon_1^i \epsilon_2^j \epsilon_3^k E_{\gamma}  I(m_{B_1^\prime},m_B,m_{B^*},q)\,, \\
\mathcal{M}_e &=& -eQ \widetilde{\beta} g_2 g_X {\epsilon}^{ijk} \epsilon_1^i \epsilon_2^j \epsilon_3^k E_{\gamma}  I(m_{B_1^\prime},m_{B^*},m_B,q)\,.
\end{eqnarray}

In the above amplitudes, the basic three-point loop function $I(q)$ is~\cite{Guo:2010ak}
\begin{eqnarray}
I(m_1,m_2,m_3,q)&=&i\int \frac{\mathrm{d}^d l}{(2\pi)^d}\frac{1}{(l^2-m_1^2+i\epsilon)[(P-l)^2-m_2^2+i\epsilon][(l-q)^2-m_3^2]+i\epsilon}\nonumber\\
&=&\frac{\mu_{12}\mu_{23}}{16\pi m_1m_2m_3}\frac{1}{\sqrt a}\left(\tan^{-1}\left(\frac{c^\prime-c}{2\sqrt{ac}}\right)+\tan^{-1}\left(\frac{2a+c^\prime-c}{2\sqrt{a(c^\prime-a)}} \right)\right).
\end{eqnarray}
Here $\mu_{ij}=m_i m_j/{(m_i+m_j)}$ are the reduced masses, $b_{12}=m_1+m_2-M$, $b_{23}=m_2+m_3+q^0-M$, and $M$ represents the mass of the initial particle.
$a=\left( \mu_{23}/{m_3} \right)^2\vec{q}^ 2$, $c=2\mu_{12}b_{12}$, and $c^\prime=2\mu_{23}b_{23}+\mu_{23}\vec{q}^2/{m_3}$.
$m_1$, $m_2$, and $m_3$ represent the masses of up, down, and right charmed mesons in the triangle loop, respectively.

The involved vector loop integral is defined as
\begin{eqnarray}
q^i I^{(1)}(m_1,m_2,m_3,q) &=& i \int \frac{\mathrm{d}^d l}{(2\pi)^d}\frac{l^i}{(l^2-m_1^2+i\epsilon)[(P-l)^2-m_2^2+i\epsilon][(l-q)^2-m_3^2]+i\epsilon}\,.
\end{eqnarray}
Using the technique of tensor reduction, we get
\begin{eqnarray}
I^{(1)}(m_1,m_2,m_3,q)\simeq \frac {\mu_{23}}{ a m_3} \left[ B(c^\prime -a)-B(c)+\frac {1} {2} (c^\prime -c) I(m_1,m_2,m_3,q) \right] \,,
\end{eqnarray}
where the function $B(c)$ is
\begin{eqnarray}
B(c) = -\frac {\mu_{12} {\mu_{23}}} {4m_1m_2m_3} \frac {\sqrt {c-i\epsilon}} {4\pi}.
\end{eqnarray}

It is worth mentioning that a factor $\sqrt {M_iM_f} m_1m_2m_3$ should be multiplied in each amplitude, when considering the nonrelativistic normalization of the bottomonium and bottomed meson fields, where $M_i$ and $M_f$ represent the masses of the initial and final particles, respectively.
\end{appendix}

\end{document}